# Malware Classification using Diluted Convolutional Neural Network with Fast Gradient Sign Method

1st Ashish Anand
*Marriott International*
Broadlands, USA
gcp.ashish2020@gmail.com

2nd Bhupendra Singh
*Marriott International*
Cary, USA
bhupendra.research1@gmail.com

3rd Sunil Khemka
*Persistent Systems*
Chicago, USA
sunilkhemka.tech@gmail.com

4th Bireswar Banerjee
*VISA*
Austin, USA
bireswar.infosys@gmail.com

5th Vishi Singh Bhatia
*Tata Consultancy Services Ltd*
Louisville, USA
vishisbhatia.research@gmail.com

6th Piyush Ranjan
*IEEE Vice Chair AeroSpace Chapter*
Edison, USA
piyush.ranjan@ieee.org

*Abstract*—Android malware has become an increasingly critical threat to organizations, society and individuals, posing significant risks to privacy, data security and infrastructure. As malware continues to evolve in terms of complexity and sophistication, the mitigation and detection of these malicious software instances have become more time consuming and challenging particularly due to the requirement of large number of features to identify potential malware. To address these challenges, this research proposes Fast Gradient Sign Method with Diluted Convolutional Neural Network (FGSM -DICNN) method for malware classification. DICNN contains diluted convolutions which increases receptive field, enabling the model to capture dispersed malware patterns across long ranges using fewer features without adding parameters. Additionally, the FGSM strategy enhance the accuracy by using one-step perturbations during training that provides more defensive advantage of lower computational cost. This integration helps to manage high classification accuracy while reducing the dependence on extensive feature sets. The proposed FGSM-DICNN model attains 99.44% accuracy while outperforming other existing approaches such as Custom Deep Neural Network (DCNN).

*Keywords—data security, diluted convolutional neural network, fast gradient sign method, malware classification, privacy.*

## I. INTRODUCTION

Malware poses a significant threat to the Internet with an average of multiple malware attacks occurring every minute as reported by an antimalware company McAfee [1]. The utilization of acquiring accessibility and infiltration of various digital resources cause undesirable outcome and increases complexity of the system [2]. The growing exposure to malware is primarily driven by an increased digitalization across domains such as e-commerce, cloud technology and convergence of Operational Technology (OT) Information Technology (IT) [3]. There are various types of malwares such as ransomware, viruses, worms and spyware each contributing to an increasing complex threat and dynamic environment [4]. The complexity and prevalence of malware attacks have escalated significantly in recent years, highlighting the severity of these cyber threats [5]. The recognition and classification of malware remains as a major challenge in research due to outdated dataset, concept shift complications and the continuous evolution of malware variants that adapt to the changing computational environment [6]. Both the growing diversity and malware proliferation has shown more problems in the cybersecurity field [7].

Dynamic execution of malware is less immune to code obfuscation problems, but this approach has other limitations. Specifically, the dynamic analysis of malware requires a precise environment to monitor and log its behavior effectively [8]. Algorithms that are utilized for malware classification are categorized into signature-based and AI-driven algorithms. Malware signatures are usually the hash values of features which are extracted by static analysis, dynamic analysis, or hybrid analysis [9]. Accurate categorization and identification of malware are critical for enhancing security in computer [10]. Highlighting the malware propagation supports security threats and eradicate threats, which calls for the requirement of more advanced and better detection tools [11]. Although advanced Deep Learning (DL) approaches assisted to recognize and categorize sophisticated malware, the evolving nature of cyber threats continues to pose significant challenges [12]. Generally, malware classification increases the posture of cybersecurity that protect integrity, confidentiality and data availability of digital system and assets, thereby minimizing cyber risks [13] [14]. While various traditional approaches demonstrated more reliable and significant outcomes when the training data amount is too high, but collecting the data can be time-consuming and has many other conditions. [15]. The contributions of this research are provided as follows:

- Diluted Convolutional Neural Network (DI-CNN) employs convolutional filters which decrease the model complexity and parameters during retaining expressive power that leads to inference with only a minor impact on accuracy and faster training.

- Fast Gradient Sign Method (FGSM) helps to generate adversarial samples during training which improves the model dependence to evasion attacks by exposing it to slightly modified malware inputs that enhances model generalizability.

The remaining part of this research is organized as follows: Literature survey provided in section 2. Workflow of proposed method is given in section 3. The results and discussion are given in 4th section. Section 5, is conclusion.

## II. LITERATURE SURVEY

In this research, the following literature review is discussed based on proposed methodology along with advantages and limitation.

Mudassar Waheed et al. [16] implemented a Random Forest (RF) model to classify among benign and malicious apps. Exploratory Data Analysis (EDA) was utilized for pre-processing to understand the significant attributes. Static and dynamic features were extracted from APK files by including permissions, intents, API calls and behavioral traces. The feature selection approaches such as wrapper, filter and embedded methods were utilized to select the most discriminative features. This method helped to minimize computational overhead while maintaining the performance. However, some sophisticated malware types were unable to bypass these checks which affected the overall accuracy of the model.

Nikolaos Polatidis et al. [17] developed a Feature Subset Selection for Android Malware Detection known as FSSDroid technique for malware detection. Mutual Information (MI) was employed to rank features based on their statistical dependencies with the class label.

The Correlation–based Feature Selection (CFS) method eliminated highly correlated features to minimize redundancy and enhanced generalizability also decreased overfitting issues. However, processing became challenging because of the utilization of same features which were available at once.

Sandeep Sharma et al. [18] introduced a Multi-Wrapper Hybrid Feature Selection Technique (MWHFST), which integrated wrapper-based feature selection for malware detection and classification. The proposed MWHFST approach was accessed across various static features and it was not limited to API calls. However, the hybrid features were often too high – dimensional features set were harnessed from android applications.

Muhammad Umar Rashid et al. [19] implemented a Custom Deep Neural Network (DCNN) that contained multiple hidden layers for hierarchies of feature learning. The dropout and batch normalization technique helped to prevent overfitting and SoftMax outcomes were utilized for classification across malware families. The Recursive Feature Elimination (RFE) technique was also used to identify the most informative features. The DCNN was useful for cybersecurity analysis for understanding threat vectors. However, the dynamic feature extraction was challenging because of slow detection in live systems.

Arvind Prasad et al. [20] developed a PermGuard framework for android malware detection which mapped permissions into exploitation and helped in incremental learning. The proposed method contained similarity based selective training approach to reduce the data required for training, and concentrate on the most relevant information which reduced the computational overhead. A PermGuard approach evaluated its dependence against adversarial attacks, thereby improving withstand attempts to bypass its detection mechanism which increased security. However, mapping was unable to capture zero-day threats.

## III. PROPOSED METHODOLOGY

In this research, the proposed FGSM -DICNN method is employed for classification of malware to enhance the classification accuracy. The missing value analysis and data cleaning is used for pre-processing to convert the raw data into formatted data. The feature selection RFE approach is used for selecting the most discriminative features to improve classification efficiency. Fig. 1 Demonstrates FGSM-DCNN proposed methodology.

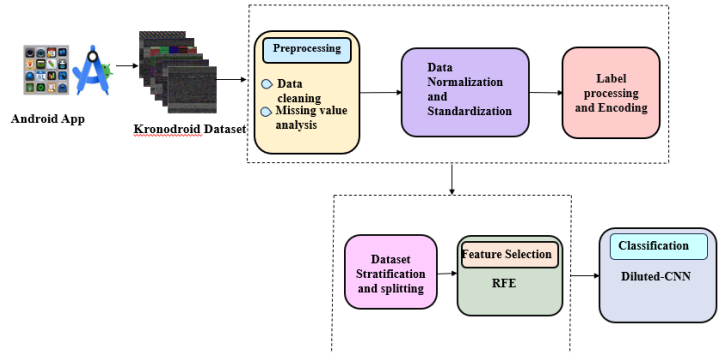

Fig. 1. Workflow for Malware classification

### A. Dataset

The KronoDroid dataset [21] comprise 78,137 malware samples, among them 41,382 are malware and 36,755 are benign with a static and dynamic of 489 features. Though there are multiple categories of malware in the database, but only three data samples such as SMS, BankBot and Airpush are selected which developed three balanced subsets.

### B. Preprocessing using Data Cleaning and Missing Value Analysis

In this section, the dataset is provided as input for preprocessing the data sample which is an essential step of data cleaning to enhance the reliability and quality of the dataset. In this phase, potential errors, inconsistencies and missing values are the primary data. This quality is assessed which revealed the dataset training and testing completeness. This comprehensive performance of missing value analysis is shown using the following formulation.

Let, this research dataset represents the $D = \{d_1, d_2, \ldots, d_n\}$, where each feature vector is $d_i$. Each feature Missing Value Ratio (MVR) is $f$ computed in Eq. (1).

$$MVR(f) = \frac{\sum_{i=1}^{n} I(d_{if} = null)}{n} \quad (1)$$

Where, indicator function is $I$ and total number of samples is $n$, the $MVR = 0$ analysis yielded across all features confirms the completeness of the dataset.

*1) Data Normalization and Standardization:* The fundamental technique like normalization which is utilized for improving numerical features denoted the model training process equally. Transforming each feature vector $x_j$ to its $x_j'$ standardized form provided in Eq. (2).

$$x_j' = \frac{x_j - \mu_j}{\sigma_j} \quad (2)$$

Where, the mean and standard deviation of features $\sigma_j$ and $\mu_j$ with respect to $j$, which is computed in Eqs. (3) and (4).

$$\mu_j = \frac{1}{n}\sum_{i=1}^{n} x_{i,j} \quad (3)$$

$$\sigma_j = \sqrt{\frac{1\sum_{i=1}^{n}(x_{i,j}-\mu_j)^2}{n}} \quad (4)$$

*2) Label Processing and Encoding:* The categorical process labels efficiently, which is encoded into numerical representation, an $y$ categorical target variable is encoded by utilizing two-phase transformation. The $Y$ is one-hot encoded matrix is defined in Eq. (5).

$$Y_{i,j} = \begin{cases} 1 \text{ if } y'_{i=j} \\ 0 \text{ otherwise} \end{cases} \quad (5)$$

Label Encoding: y → y where $y \in \{0,1,\ldots,k-1\}$ for $k$ classes.

One-Hot Encoding: y→ Y where $Y \in \{0,1\}^{n \times k}$.

*3) Dataset Stratification and Splitting:* Stratified sampling is used to ensure balanced representation of classes in the training and validation sets. The partitioning of the dataset is done while managing class distribution. For every $c$ class, the below mentioned constraints are handled from Eqs. (6), (7) and (8):

$$\frac{|S_c^{train}|}{|S^{train}|} \approx \frac{|S_c^{val}|}{|S^{val}|} \approx P_c \quad (6)$$

$$|S^{val}| = \eta|s| \quad (7)$$

$$|S^{train}| = (1-\eta)|S| \quad (8)$$

Where, subset of samples $S_c$ belongs $g$ to class $c$, and original class propagation in the data is $S_c$. The ratio of splitting is $\eta$ in a set of 0.2 respectively.

## C. Feature Selection using RFE

In the malware classification task, features like API calls, extracted attributes, and header information which assembled into a model-specific feature set. The RFE approach systematically removes the least significant features in each iteration based on their performance scores, while retraining the model. This process is repeated until a target number of optimal predictive features is obtained. This RFE decreases overfitting by improving interpretability and increase computational efficiency while managing high-dimensional malware data. Irrelevant attributes are eliminated by focusing on the most discriminative pattern distinguishing benign and malicious software. This feature subset not only improves classification performance but also enables faster, more reliable malware detection.

## D. Classification

The proposed DI-CNN approach helped to classify the malicious samples for an accurate differentiation. The baseline CNN has a limitation of convolutional kernel size representing main issues, initially CNN captured only short-term dependencies and then secondly the convolutional kernels were expanded which increased significant parameter.

DI-CNN enhanced the overall network efficiency while improving the collection of larger amounts of data and fields size. This classification efficiently enhanced the extended receptive field in the task of classification. The DI-CNN one-way extension "1" integrated the "$E$" signal with the "R" kernel that contained "$r$" particular field size. This can be extended up to 2Dimensional convolution given in Eq. (9).

$$(E * p_x R)_s = \sum_\tau R_\tau E_{s-p_x\tau} \quad (9)$$

Where, increased dilation width exponentially is $E * p_x$ of "l" corresponding layer in the dilated convolution $x$ respectively.

*1) Fast Gradient Sign Method:* This input perturbation technique takes gradient loss which concerns gradient direction, while adding small perturbation and input data expressed in Eq. (10).

$$x_{adv=}X + \epsilon.sign(\nabla_x J(\theta,x,y)) \quad (10)$$

Where the perturbation size determination is Epsilon ($\epsilon$) - 0.01, 0.05, 0.1 and $x$ are typical values in the input representing the strength of perturbation and the true class label is $y$, and the loss function is $J$, these are generated adversarial samples which can be used for malware classification.

## IV. EXPERIMENT RESULT AND DISCUSSION

The performance of the developed FGSM-DICNN is determined based on the dataset which is used for training and testing. The developed FGSM-DICNN is implemented in python 3.8 environment and the system run with Intel Core i7-4200U CPU processor, Windows 10, 64-bit OS and 16GB RAM. The developed FGSM-DICNN technique is estimated using various analysis metrics, namely, recall, precision, F1-Score and precision. The mathematical expressions for each metric are formulated from Eqs. (11) – (14), respectively.

The percentage of correctly classified pixels in the dataset is called as accuracy.

$$Accuracy = \frac{TP}{TP+FN+TN+FP} \quad (11)$$

The percentage of true positive values among all predicted values to be positive is called precision.

$$Precision = \frac{TP}{TP+FP} \quad (12)$$

The fraction calculation of all positive samples estimated to be positive using the predictive utility equation of recall.

$$Recall = \frac{TP}{TP+FN} \quad (13)$$

The F1-Score measures the reciprocity between precision and recall, as shown below.

$$F1 = \frac{2 \times Precision \times Recall}{Precision \times Recall} \quad (14)$$

Where, $TP$ – True Positive; $TN$ – True Negative; $FP$ – False Positive; $FN$ – False Negative.

## A. Performance Analysis

The comprehensive analysis of the developed FGSM-DICNN is evaluated while utilizing KronoDroid database. An efficiency of the designed approach is validated in this section by using the performance of various metrics. Fig. 2 represents the analysis of malware classification.

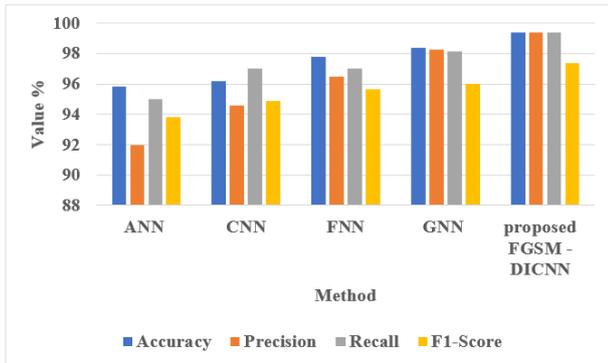

Fig. 2. Analysis of FGSM -DICNN with existing DL model

In Fig. 2 Analysis of FGSM-DICNN with existing DL model database. The efficiency of the different DL methods Artificial Neural Network [ANN], Convolution Neural network [CNN], and Feed Forward Lyer [FNN] and Graph Neural Network [GNN] is compared with the developed FGSM-DICNN approach. The comparison with these methods showed that the developed FGSM-DICNN had achieved the best capability of exploration and addressed the types of malwares which led to accurate classification. The FGSM-DICNN method attains an accuracy of 99.41% for KronoDroid dataset.

## B. Comparative Analysis

The proposed FGSM-DICNN approach comparison is presented by utilizing many performance metrics. Table 1 shows the comparison outcomes of the developed approach with traditional methods namely RF [16], FSSDroid [17], MWHFST [18], DDCNN [19] and PermGuard [20] respectively. As compared to existing models, FGSM -DICNN has led to efficient accuracy results. The method FGSM -DICNN performance is evaluated by utilizing evaluation metrics.

TABLE I. COMPARATIVE ANALYSIS FOR AMONG EXISTING APPROACHES AND PROPOSED METHODOLOGY

| Method | Accuracy | Precision | Recall | F1-Score |
|---|---|---|---|---|
| RF [16] | 87.24 | 90.14 | 74.20 | 81.39 |
| FSSDroid [17] | 98.50 | 98.50 | 98.60 | - |
| MWHFST [18] | 88 | 88 | 88 | 88 |
| DCNN [19] | 94 | 94 | 94 | 94 |
| PermGuard [20] | 99.33 | 99.34 | 99.32 | 99.33 |
| Proposed FGSM -DICNN | 99.41 | 99.38 | 99. 37 | 99.36 |

In Table 1, the proposed methodology FGSM -DICNN is introduced along with the performance metrics of KronoDroid database. The Proposed method performs efficiently compared to analyzed methods to enhance the detection accuracy. The FGSM -DICNN method attains 99.41% of accuracy for KronoDroid dataset.

## C. Discussion

Android operating system is not just the most commonly employed mobile operating system, but also the most lucrative target for cybercriminals due to its extensive user base. The dilated convolutional layer is used for classification of android malware which enables the model to capture pattern across distributed segments and capture long range dependencies of features and feature representation in application. Thereby using dilation, a model enlarges its receptive fields without sacrificing resolution that is mainly valuable for malicious dispersed detection. The FGSM approach is provided perturbed samples while training which it is difficult for the model to solve that mislead detection and subtle alter inputs. This integration improved the classification accuracy by differentiating malicious and benign samples. The proposed FGSM-DICNN model attained an accuracy of 99.41% for KronoDroid dataset when compared with other methods.

## V. CONCLUSION

Android is the most widely used mobile operating system which is responsible for maintaining a wide variety of data from simple message to sensitive banking details. The malware targeted explosive usage of this platform which has made it imperative to adopt DICNN for efficient classification and detection of malware. The proposed FGSM-DICNN method helped to enhance the model ability to distinguish patterns that is spread across feature space distant regions by identifying problems. An FGSM approach improve the adversarial dependence via lightweight effective techniques which has an exposure to perturbed samples. This integration is used to significantly classify the malware in android application. The proposed FGSM-DICNN model attained an accuracy of 99.41% for KronoDroid dataset when compared to other methods. In future years, work is done using an advanced DL models to defend against non-defensive and non-dilated attacks to improve defensive strength and context modelling, thereby making it more applicable for real-world malware system.

REFERENCES

[1] A. Bensaoud and J. Kalita, "CNN-LSTM and transfer learning models for malware classification based on opcodes and API calls," Knowl.-Based Syst., vol. 290, p. 111543, April 2024.

[2] V. Vouvoutsis, F. Casino, and C. Patsakis, "Beyond the sandbox: Leveraging symbolic execution for evasive malware classification," Comput. Secur., vol. 149, p. 104193, February 2025.

[3] S. J. I. Ismail, B. Rahardjo, T. Juhana, and Y. Musashi, "MalSSL—Self-supervised learning for accurate and label-efficient malware classification," IEEE Access, vol. 12, pp. 58823–58835, April 2024.

[4] G. Kale, G. E. Bostancı, and F. V. Celebi, "Evolutionary feature selection for machine learning based malware classification," Eng. Sci. Technol. Int. J., vol. 56, p. 101762, August 2024.

[5] S. Duraibi, "Enhanced image-based malware classification using snake optimization algorithm with deep convolutional neural network," IEEE Access, vol. 12, pp. 95047-95057, July 2024.

[6] R. Alguliyev, R. Aliguliyev, and L. Sukhostat, "Radon transform based malware classification in cyber-physical system using deep learning," Results Control Optim., vol. 14, p. 100382, March 2024.

[7] M. E. Farfoura, I. Mashal, A. Alkhatib, R. M. Batyha, and D. Rosiyadi, "A novel lightweight machine learning framework for IoT malware classification based on matrix block mean downsampling," Ain Shams Eng. J., vol. 16, p. 103205, January 2025.

[8] D. Gibert, G. Zizzo, Q. Le, and J. Planes, "Adversarial robustness of deep learning-based malware detectors via (de) randomized smoothing," IEEE Access, vol. 12, pp. 61152–61162, April 2024.

[9] A. Egitmen, A. G. Yavuz, and S. Yavuz, "TRConv: Multi-platform malware classification via target regulated convolutions," IEEE Access, vol. 12, pp. 71492–71504, May 2024.

[10] K. Alfarsi, S. Rasheed, and I. Ahmad, "Malware classification using few-shot learning approach," Inf., vol. 15, p. 722, November 2024.

[11] D. Z. Syeda and M. N. Asghar, "Dynamic malware classification and API categorisation of Windows portable executable files using machine learning," Appl. Sci., vol. 14, p. 1015, January 2024.

[12] M. Ashawa, N. Owoh, S. Hosseinzadeh, and J. Osamor, "Enhanced image-based malware classification using transformer-based convolutional neural networks (CNNs)," Electron., vol. 13, p. 4081, October 2024.

[13] O. Sharma, A. Sharma, and A. Kalia, "MIGAN: GAN for facilitating malware image synthesis with improved malware classification on novel dataset," Expert Syst. Appl., vol. 241, p. 122678, May 2024.

[14] J. Jeon, B. Jeong, S. Baek, and Y. S. Jeong, "TMaD: Three-tier malware detection using multi-view feature for secure convergence ICT environments," Expert Syst., vol. 42, p. e13684, July 2025.

[15] P. Wu, M. Gao, F. Sun, X. Wang, and L. Pan, "Multi-perspective API call sequence behavior analysis and fusion for malware classification," Comput. Secur., vol. 148, p. 104177, January 2025.

[16] M. Waheed and S. Qadir, "Effective and efficient Android malware detection and category classification using the enhanced Kronodroid dataset," Secur. Commun. Netw., vol. 2024, p. 7382302, April 2024.

[17] N. Polatidis, S. Kapetanakis, M. Trovati, I. Korkontzelos, and Y. Manolopoulos, "FSSDroid: Feature subset selection for Android malware detection," World Wide Web, vol. 27, p. 50, July 2024.

[18] S. Sharma, R. Chhikara, and K. Khanna, "A novel feature selection technique: Detection and classification of Android malware," Egypt. Inform. J., vol. 29, p. 100618, March 2025.

[19] M. U. Rashid, S. Qureshi, A. Abid, S. S. Alqahtany, A. Alqazzaz, M. S. Al Reshan, and A. Shaikh, "Hybrid Android malware detection and classification using deep neural networks," Int. J. Comput. Intell. Syst., vol. 18, p. 52, March 2025.

[20] A. Prasad, S. Chandra, M. Uddin, T. Al-Shehari, N. A. Alsadhan, and S. S. Ullah, "PermGuard: A scalable framework for Android malware detection using permission-to-exploitation mapping," IEEE Access, vol. 13, pp. 507-528, December 2024.

[21] KronoDroid dataset: https://www.kaggle.com/datasets/dhoogla/kronodroid-2021 (Accessed on September 2025).